\RequirePackage{lineno}
\documentclass[aps,prd,twocolumn,showpacs,superscriptaddress,groupedaddress,floatfix]{revtex4}  
\usepackage{graphicx}
\usepackage{dcolumn}
\usepackage{bm}
\usepackage{amssymb}
\usepackage{amsmath}
\usepackage{xspace}
\usepackage{color}
\usepackage{enumerate}
\usepackage{afterpage}
\newcommand{\ppbar}{$p\bar{p}$~}

\newcommand{\pt}{${p}_{T}$\xspace}

\newcommand{\Ptg}{p_{T}^{\gamma}}
\newcommand{\Ptj}{p_{T}^\text{jet}}

\newcommand{\ptgamone}{${p}^{\gamma_{1}}_{\rm T}$\xspace}
\newcommand{\ptgamtwo}{${p}^{\gamma_{2}}_{\rm T}$\xspace}

\newcommand{\la}{\langle}
\newcommand{\ra}{\rangle}
\newcommand{\lt}{\!<\!}
\newcommand{\gt}{\!>\!}

\newcommand{\sigmaeff}{$\sigma_{\rm eff}$}

\newcommand{\gamgam}{$\gamma\gamma$\xspace}
\newcommand{\ggjj}{$\gamma\gamma+{\rm dijet}$\xspace}
\newcommand{\gpj}{$\gamma+{\rm jet}$\xspace}

\newcommand{\dS}{$\Delta S$~}

\newcommand{\ndp}{$N_{\mathrm{DP}}$\xspace}
\newcommand{\ndi}{$N_{\mathrm{DI}}$\xspace}
\newcommand{\fdp}{$f_{\mathrm{DP}}$\xspace}
\newcommand{\fdi}{$f_{\mathrm{DI}}$\xspace}

\newcommand{\onevtx}{\normalfont {1VTX}\xspace}
\newcommand{\twovtx}{\normalfont {2VTX}\xspace}

\newcommand{\mixdp}{\normalfont{MIXDP}\xspace}
\newcommand{\mixdi}{\normalfont{MIXDPI}\xspace}
\newcommand{\mcdp}{\normalfont{MCDP}\xspace}
\newcommand{\mcdi}{\normalfont{MCDI}\xspace}

\newcommand{\sponevtx}{\normalfont{SP1VTX}\xspace}

\newcommand{\pythia}{$\textsc{pythia}$\xspace}
\newcommand{\sherpa}{$\textsc{sherpa}$\xspace}

\newcommand{\GeV}{\ensuremath{\text{GeV}}\xspace}

\begin{document}


\widetext
\hspace{5.2in} \mbox{FERMILAB-PUB-15-562-E}

\title{ 
Study of double parton interactions in diphoton + dijet events in $p\bar{p}$ collisions 
at $\sqrt{s} = 1.96$~TeV}
%
\affiliation{LAFEX, Centro Brasileiro de Pesquisas F\'{i}sicas, Rio de Janeiro, Brazil}
\affiliation{Universidade do Estado do Rio de Janeiro, Rio de Janeiro, Brazil}
\affiliation{Universidade Federal do ABC, Santo Andr\'e, Brazil}
\affiliation{University of Science and Technology of China, Hefei, People's Republic of China}
\affiliation{Universidad de los Andes, Bogot\'a, Colombia}
\affiliation{Charles University, Faculty of Mathematics and Physics, Center for Particle Physics, Prague, Czech Republic}
\affiliation{Czech Technical University in Prague, Prague, Czech Republic}
\affiliation{Institute of Physics, Academy of Sciences of the Czech Republic, Prague, Czech Republic}
\affiliation{Universidad San Francisco de Quito, Quito, Ecuador}
\affiliation{LPC, Universit\'e Blaise Pascal, CNRS/IN2P3, Clermont, France}
\affiliation{LPSC, Universit\'e Joseph Fourier Grenoble 1, CNRS/IN2P3, Institut National Polytechnique de Grenoble, Grenoble, France}
\affiliation{CPPM, Aix-Marseille Universit\'e, CNRS/IN2P3, Marseille, France}
\affiliation{LAL, Univ. Paris-Sud, CNRS/IN2P3, Universit\'e Paris-Saclay, Orsay, France}
\affiliation{LPNHE, Universit\'es Paris VI and VII, CNRS/IN2P3, Paris, France}
\affiliation{CEA, Irfu, SPP, Saclay, France}
\affiliation{IPHC, Universit\'e de Strasbourg, CNRS/IN2P3, Strasbourg, France}
\affiliation{IPNL, Universit\'e Lyon 1, CNRS/IN2P3, Villeurbanne, France and Universit\'e de Lyon, Lyon, France}
\affiliation{III. Physikalisches Institut A, RWTH Aachen University, Aachen, Germany}
\affiliation{Physikalisches Institut, Universit\"at Freiburg, Freiburg, Germany}
\affiliation{II. Physikalisches Institut, Georg-August-Universit\"at G\"ottingen, G\"ottingen, Germany}
\affiliation{Institut f\"ur Physik, Universit\"at Mainz, Mainz, Germany}
\affiliation{Ludwig-Maximilians-Universit\"at M\"unchen, M\"unchen, Germany}
\affiliation{Panjab University, Chandigarh, India}
\affiliation{Delhi University, Delhi, India}
\affiliation{Tata Institute of Fundamental Research, Mumbai, India}
\affiliation{University College Dublin, Dublin, Ireland}
\affiliation{Korea Detector Laboratory, Korea University, Seoul, Korea}
\affiliation{CINVESTAV, Mexico City, Mexico}
\affiliation{Nikhef, Science Park, Amsterdam, the Netherlands}
\affiliation{Radboud University Nijmegen, Nijmegen, the Netherlands}
\affiliation{Joint Institute for Nuclear Research, Dubna, Russia}
\affiliation{Institute for Theoretical and Experimental Physics, Moscow, Russia}
\affiliation{Moscow State University, Moscow, Russia}
\affiliation{Institute for High Energy Physics, Protvino, Russia}
\affiliation{Petersburg Nuclear Physics Institute, St. Petersburg, Russia}
\affiliation{Instituci\'{o} Catalana de Recerca i Estudis Avan\c{c}ats (ICREA) and Institut de F\'{i}sica d'Altes Energies (IFAE), Barcelona, Spain}
\affiliation{Uppsala University, Uppsala, Sweden}
\affiliation{Taras Shevchenko National University of Kyiv, Kiev, Ukraine}
\affiliation{Lancaster University, Lancaster LA1 4YB, United Kingdom}
\affiliation{Imperial College London, London SW7 2AZ, United Kingdom}
\affiliation{The University of Manchester, Manchester M13 9PL, United Kingdom}
\affiliation{University of Arizona, Tucson, Arizona 85721, USA}
\affiliation{University of California Riverside, Riverside, California 92521, USA}
\affiliation{Florida State University, Tallahassee, Florida 32306, USA}
\affiliation{Fermi National Accelerator Laboratory, Batavia, Illinois 60510, USA}
\affiliation{University of Illinois at Chicago, Chicago, Illinois 60607, USA}
\affiliation{Northern Illinois University, DeKalb, Illinois 60115, USA}
\affiliation{Northwestern University, Evanston, Illinois 60208, USA}
\affiliation{Indiana University, Bloomington, Indiana 47405, USA}
\affiliation{Purdue University Calumet, Hammond, Indiana 46323, USA}
\affiliation{University of Notre Dame, Notre Dame, Indiana 46556, USA}
\affiliation{Iowa State University, Ames, Iowa 50011, USA}
\affiliation{University of Kansas, Lawrence, Kansas 66045, USA}
\affiliation{Louisiana Tech University, Ruston, Louisiana 71272, USA}
\affiliation{Northeastern University, Boston, Massachusetts 02115, USA}
\affiliation{University of Michigan, Ann Arbor, Michigan 48109, USA}
\affiliation{Michigan State University, East Lansing, Michigan 48824, USA}
\affiliation{University of Mississippi, University, Mississippi 38677, USA}
\affiliation{University of Nebraska, Lincoln, Nebraska 68588, USA}
\affiliation{Rutgers University, Piscataway, New Jersey 08855, USA}
\affiliation{Princeton University, Princeton, New Jersey 08544, USA}
\affiliation{State University of New York, Buffalo, New York 14260, USA}
\affiliation{University of Rochester, Rochester, New York 14627, USA}
\affiliation{State University of New York, Stony Brook, New York 11794, USA}
\affiliation{Brookhaven National Laboratory, Upton, New York 11973, USA}
\affiliation{Langston University, Langston, Oklahoma 73050, USA}
\affiliation{University of Oklahoma, Norman, Oklahoma 73019, USA}
\affiliation{Oklahoma State University, Stillwater, Oklahoma 74078, USA}
\affiliation{Oregon State University, Corvallis, Oregon 97331, USA}
\affiliation{Brown University, Providence, Rhode Island 02912, USA}
\affiliation{University of Texas, Arlington, Texas 76019, USA}
\affiliation{Southern Methodist University, Dallas, Texas 75275, USA}
\affiliation{Rice University, Houston, Texas 77005, USA}
\affiliation{University of Virginia, Charlottesville, Virginia 22904, USA}
\affiliation{University of Washington, Seattle, Washington 98195, USA}
\author{V.M.~Abazov} \affiliation{Joint Institute for Nuclear Research, Dubna, Russia}
\author{B.~Abbott} \affiliation{University of Oklahoma, Norman, Oklahoma 73019, USA}
\author{B.S.~Acharya} \affiliation{Tata Institute of Fundamental Research, Mumbai, India}
\author{M.~Adams} \affiliation{University of Illinois at Chicago, Chicago, Illinois 60607, USA}
\author{T.~Adams} \affiliation{Florida State University, Tallahassee, Florida 32306, USA}
\author{J.P.~Agnew} \affiliation{The University of Manchester, Manchester M13 9PL, United Kingdom}
\author{G.D.~Alexeev} \affiliation{Joint Institute for Nuclear Research, Dubna, Russia}
\author{G.~Alkhazov} \affiliation{Petersburg Nuclear Physics Institute, St. Petersburg, Russia}
\author{A.~Alton$^{a}$} \affiliation{University of Michigan, Ann Arbor, Michigan 48109, USA}
\author{A.~Askew} \affiliation{Florida State University, Tallahassee, Florida 32306, USA}
% ...
%
%
\collaboration{The D0 Collaboration\footnote{with visitors from
$^{a}$Augustana College, Sioux Falls, SD, USA,
$^{b}$The University of Liverpool, Liverpool, UK,
$^{c}$DESY, Hamburg, Germany,
$^{d}$CONACyT, Mexico City, Mexico,
$^{e}$SLAC, Menlo Park, CA, USA,
$^{f}$University College London, London, UK,
$^{g}$Centro de Investigacion en Computacion - IPN, Mexico City, Mexico,
$^{h}$Universidade Estadual Paulista, S\~ao Paulo, Brazil,
$^{i}$Karlsruher Institut f\"ur Technologie (KIT) - Steinbuch Centre for Computing (SCC),
D-76128 Karlsruhe, Germany,
$^{j}$Office of Science, U.S. Department of Energy, Washington, D.C. 20585, USA,
$^{k}$American Association for the Advancement of Science, Washington, D.C. 20005, USA,
$^{l}$Kiev Institute for Nuclear Research, Kiev, Ukraine,
$^{m}$University of Maryland, College Park, MD 20742, USA,
$^{n}$European Orgnaization for Nuclear Research (CERN), Geneva, Switzerland
and
$^{o}$Purdue University, West Lafayette, IN 47907, USA.
$^{\ddag}$Deceased.
}} \noaffiliation
\vskip 0.25cm

\date{March 17, 2016}

\begin{abstract}
We use a sample of diphoton + dijet events
to measure the effective cross section of double parton interactions, which characterizes the area containing the interacting partons in proton-antiproton collisions,
 and find it to be  $\sigma_{\rm eff} = 19.3$~$\pm$~$1.4({\rm stat})$~$\pm$~$7.8({\rm syst})$ mb.
The sample was collected by the D0 detector at the Fermilab Tevatron collider in $p\bar{p}$ collisions at $\sqrt{s} = 1.96$ TeV and corresponds to an integrated luminosity of 8.7 fb$^{-1}$.
\end{abstract}


\maketitle

\section{Introduction}
\label{Sec:Intro}

Many features of high energy inelastic hadron collisions
are directly dependent on the parton structure of hadrons, which
is not yet completely understood either at the theoretical or
experimental levels.
Studies of this structure generally rely on
a theoretical model of
inelastic scattering of high energy nucleons,
where a single parton (quark or gluon from one nucleon or a lepton in Deep Inelastic Scattering (DIS) experiments)
interacts with a single parton from another nucleon.
In this approach, the other
``spectator'' partons which do not take part in a hard
$2 \to 2$ parton collision are included in the so-called
``underlying event.''

Information regarding the abundance of simultaneous double parton (DP)
interactions comprising two separate hard parton scatterings within a single hadron-hadron collision~\cite{Landsh,Goebel,TH1,TH1b,TH2,TH2b,TH3,Threl,Threlb,Threl2,Flen,Sjost,Snigir,Snigirb,Strikman,Strikmanb} 
is a subject of great interest,
because the growing LHC luminosity provides an opportunity to 
search for signals from new physics for
which the DP events constitute a significant background, especially
in the multijet final state. For example, processes such as the associated
production of the Higgs and $W$ bosons, with the Higgs boson decaying into a pair of $b$ quarks, have substantial DP backgrounds~\cite{Bandurin:2010gn}.

Several relevant measurements have been already performed using hadron collisions at
$\sqrt{s}=63$ GeV \cite{AFS}, $\sqrt{s}=630$ GeV \cite{UA2},
$\sqrt{s}=1.8$ TeV \cite{CDF93,CDF97}, $\sqrt{s}=1.96$ TeV \cite{D02010,D02011,D02014,D0jj,D0jy},
$\sqrt{s}=7$ TeV \cite{ATLAS2013, CMS2013,LHCbD0,LHCbj}, and $\sqrt{s}=8$ TeV \cite{LHCbD0}.
The first three measurements utilize
a four jet
final state, where the transverse momentum of the jets in each jet pair is balanced,
resulting in the jets produced at almost opposite azimuthal angles. 
AFS~\cite{AFS} has found (for jet transverse energy $E^{\rm jet}_{T} > 4$ GeV and pseudorapidity~\cite{d0_coordinate}
$|\eta^{\rm jet} | \leq 1$) the ratio of DP/2jet cross sections to be
$6\% \pm 1.5\%$({\rm stat}) $\pm$ 2.2\%({\rm syst}).
UA2 \cite{UA2} retained only jet clusters
with transverse momentum $p_{T}^{\rm jet} > 15$ GeV
and $|\eta^{\rm jet}| < 2$ and set a 95\% C.L. limit on the value
of the DP cross section, $\sigma_{\rm DP} \leq 0.82$ nb.
The CDF measurement of the DP fraction in four jet events~\cite{CDF93} found a DP cross section of
  $\sigma_{\rm DP}= 63^{+32}_{-28}$ nb
for jets having
$p_{T}^{\rm jet} \geq 25$ GeV and $|\eta^{\rm jet}| \leq 3.5$.
Additional CDF and D0 measurements \cite{CDF97,D02010,D02011,D02014} are based on
the DP process comprising
two parton scatterings with one of them having a dijet
final state and the other having a $\gamma+$jet or $\gamma+b(c)$-jet final state.
D0 and LHCb measurements \cite{D0jj,D0jy,LHCbD0,LHCbj} probe the final states containing heavy quarkonia.
 In Refs.~\cite{D0jy,LHCbD0}, the production of the studied final states
  in DP scattering is predicted to dominate
the production in a single parton (SP) scattering.
In this paper, we report the first measurement of DP scattering in the diphoton-dijet ($\gamma \gamma$ +jj) channel.

As shown experimentally
in Refs.~\cite{CDF93, CDF97,D02010} and described in Ref.~\cite{Han},
the substitution of one of the two
dijet parton processes by a photon jet or a diphoton process leads to
about an order of magnitude increase in the ratio of the 
DP cross section to the cross section of the SP scattering
for the production of the same final state.
This improves the ability to characterize the DP contribution in the data.
 Additionally, a technique for
extracting an important physical parameter,
$\sigma_{\rm eff}$, has been proposed in Ref.~\cite{CDF97}. This method uses only quantities
obtained from data analysis and minimizes
theoretical assumptions that were used in the previous measurements.

The parameter, $\sigma_{\rm eff}$, is related to the distance
between partons in the nucleon \cite{TH1, TH1b,TH2, TH2b,Threl, Threlb, AFS, UA2, CDF93, CDF97},
\begin{eqnarray}
\sigma_{\rm eff}^{-1} = \int d^2\beta [F({\bf \beta})]^2
\end{eqnarray}
with $F({\bf \beta})=\int f({\bf b}) f({\bf b}-{\bf \beta})d^2b$,
where ${\bf \beta}$ is the vector impact parameter of the two colliding hadrons
and $f({\bf b})$ is a function describing the transverse spatial distribution
of the partonic matter inside a hadron \cite{Threl,Threlb,Threl2}.
The $f({\bf b})$ may depend on the parton flavor.

The cross section for double parton scattering, $\sigma_{\rm DP}$, is related to $\sigma_{\rm eff}$~\cite{UA2, CDF93, CDF97} 
for the 2-$\gamma$ and 2-jet process as
\begin{eqnarray}
\sigma_{\rm DP} \equiv
 \frac{m}{2} \frac{\sigma^{\gamma \gamma} \sigma^{jj}}{\sigma_{\rm eff} }.
\label{eq:sigmaeff_main}
\end{eqnarray}
The factor of $1/2$ is due to the assumption that the probability
of multiple parton interactions inside the proton follows a
Poisson distribution~\cite {TH3}. For this analysis, the factor $m$ is equal to
$2$ because the diphoton and double jet production
processes are distinguishable (in the case of 4-jet production,
i.e. two dijet processes, $m=1$).
Table \ref{tab:sigeff_world} summarizes the available data on the measurements of $\sigma_{\rm eff}$.
\begin{table*}[htpb]
\vskip1mm
\begin{center}
\caption{Summary of the results, experimental parameters, and event selection criteria for the double parton analyses 
performed by the AFS, UA2, CDF, D0, ATLAS, CMS, and LHCb Collaborations 
(no uncertainties are available for the AFS result).}
\label{tab:sigeff_world}
\vskip 1mm
\begin{tabular}{|c||cclll|} \hline
$  $ & $ \sqrt{s} $ (GeV) & Final state & $p_{T}^{min}$ (GeV/c) & $\eta $ range & Result \\\hline
 AFS, 1986 \cite{AFS} & $ 63 $  & $ 4$ jets       & $\Ptj>4$ & $|\eta^{\rm jet}|<1$  & $ \sigma_{\rm eff}$ $\sim 5 $ mb \\\hline
 UA2, 1991 \cite{UA2} & $ 630 $  & $ 4$ jets       & $\Ptj>15$ & $|\eta^{\rm jet}|<2$  & $ \sigma_{\rm eff} > 8.3$ mb (95\% C.L.) \\\hline 
 CDF, 1993 \cite{CDF93} & $ 1800 $ & $ 4$ jets       & $\Ptj>25$ & $|\eta^{\rm jet}|<3.5$ & $ \sigma_{\rm eff} = 12.1_{-5.4}^{+10.7}$ mb \\\hline
 D0, 2014 \cite{D0jj} & $ 1960 $ & $ J/\psi J/\psi$ & $p^{J/\psi}_T>4$ & $|\eta^{J/\psi}|<2.2$ & $\sigma_{\rm eff} = 4.8 \pm 0.5 \pm 2.5$ mb \\\hline
 LHCb, 2015 \cite{LHCbD0} & $ 7000, 8000$ & $ \Upsilon D^{0+}$ & $p^\Upsilon_T<15$ & $2.0<y^\Upsilon<4.5$ & $\sigma_{\rm eff} = 18 \pm 1.8$ mb \\\hline
 D0, 2015 \cite{D0jy} & $ 1960 $ & $ J/\psi \Upsilon$ & $p^{\mu}_T>2$ & $|\eta^{\mu}|<2$ & $\sigma_{\rm eff} = 2.2 \pm 0.7 \pm 0.9$ mb \\\hline
 CDF, 1997 \cite{CDF97} & $ 1800 $ & $\gamma + 3$ jets   & $\Ptj>6$  & $|\eta^{\rm jet}|<3.5$ &                        \\
      &      &           & $\Ptg>16$ & $|\eta^{\gamma}|<0.9$ & $ \sigma_{\rm eff} = 14.5$$\pm$$1.7_{-2.3}^{+1.7}$ mb\\\hline
 D0, 2009 \cite{D02010} & $ 1960 $ & $\gamma + 3$ jets   & $60<\Ptg<80$&$|\eta^{\gamma}|<1.0$ & $ \sigma_{\rm eff} = 16.4 \pm 2.3$ mb \\
      &      &           &       &$1.5<|\eta^{\gamma}|<2.5$ &                     \\\hline
 D0, 2014 \cite{D02014} & $ 1960 $ & $\gamma + 3$ jets   & $\Ptg>26$&$|\eta^{\gamma}|<1.0$ & $ \sigma_{\rm eff} = 12.7\pm1.3$ mb \\
      &      &           &       &$1.5<|\eta^{\gamma}|<2.5$ &                     \\\hline
 D0, 2014 \cite{D02014} & $ 1960 $ & $\gamma + b/c$ jet + 2 jets  & $\Ptg>26$&$|\eta^{\gamma}|<1.0$ & $ \sigma_{\rm eff} = 14.6\pm3.3$ mb \\
      &      &           &       &$1.5<|\eta^{\gamma}|<2.5$ &                     \\\hline
 ATLAS, 2013 \cite{ATLAS2013} & $ 7000 $ & $W + 2$ jets     & $\Ptj>20$& $|\eta^{\rm jet}|<2.8$ & $ \sigma_{\rm eff} = 15$$\pm$$3_{-3}^{+5}$ mb\\\hline
 CMS, 2014 \cite{CMS2013} & $ 7000 $ & $W + 2$ jets     & $\Ptj>20$& $|\eta^{\rm jet}|<2.0$ & $ \sigma_{\rm eff} = 20.7\pm6.6$ mb\\\hline
\end{tabular}
\end{center}
\vskip 2mm
\end{table*}
The goal of this study is to obtain the DP rate and the effective cross section
in the diphoton+dijet final state.

\begin{figure}[h]
\vspace*{0mm}
\hspace*{5mm} \includegraphics[scale=0.33]{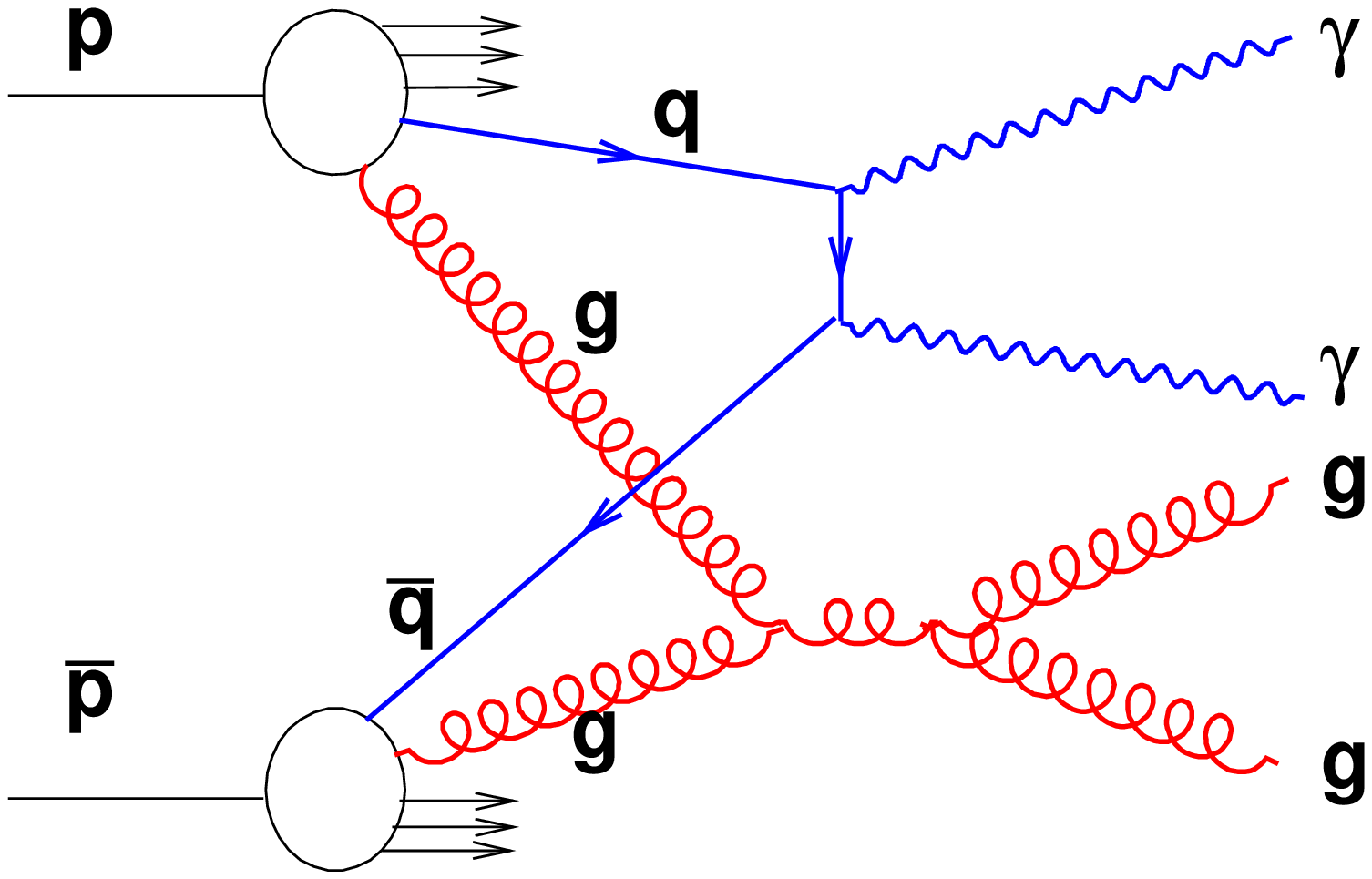}
\hspace*{5mm} \includegraphics[scale=0.33]{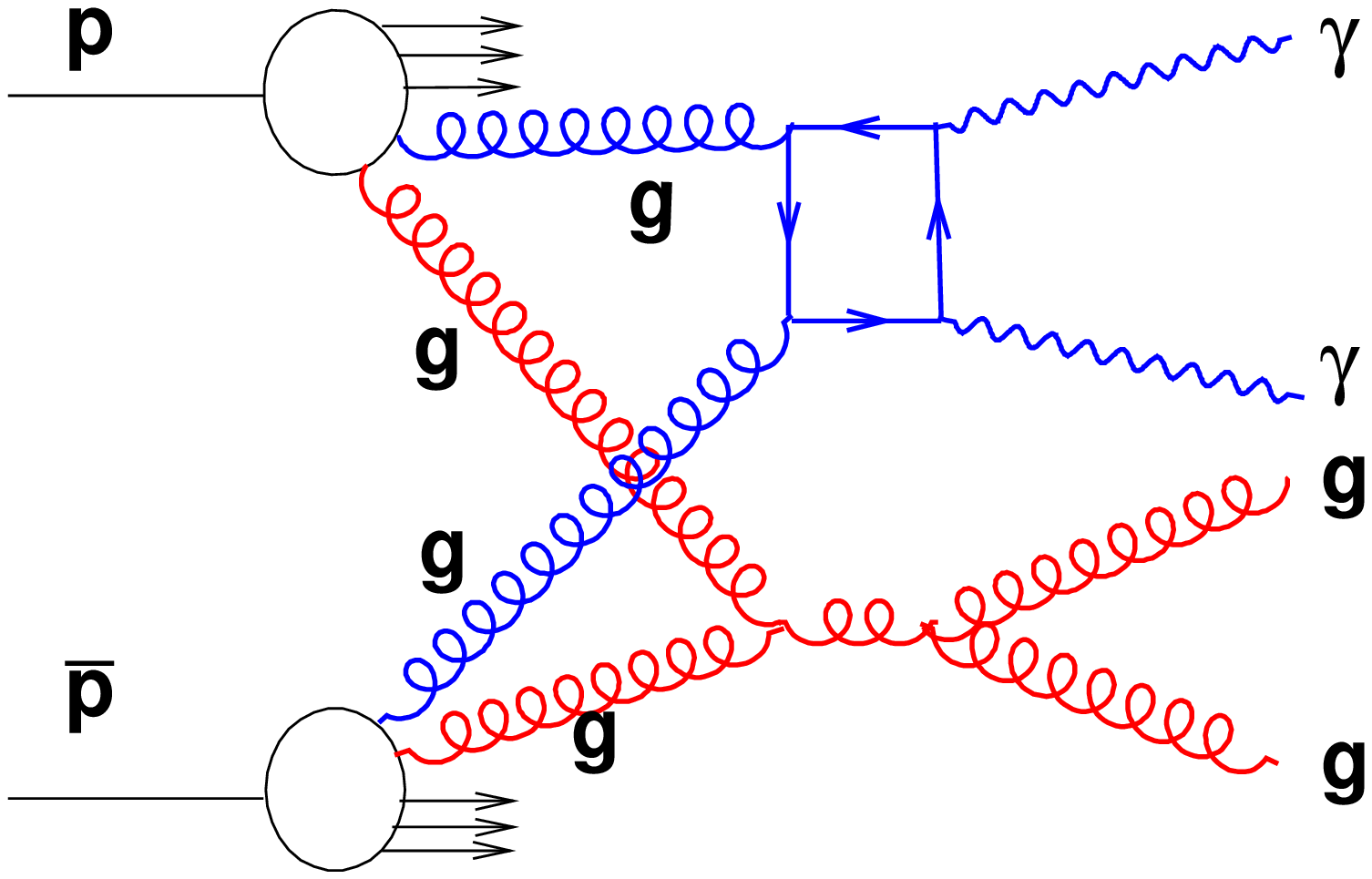}
\vspace*{-2mm}
\caption{ Schematic view of DP scattering processes producing \ggjj final state. The $\gamma \gamma$ process
is shown for the $q \bar q$ scattering (above, light, blue online) and box $gg$ diagram (below, light, blue online). The additional dijet scattering is a darker diagram (red online).}
\label{fig:DP}
\vspace*{5mm}
\end{figure}

The main contributions to diphoton production at the Tevatron are from the
$q\bar{q} \to \gamma\gamma$ and $gg \to \gamma\gamma$ via direct $2 \rightarrow 2$ partonic processes,
as well as from bremsstrahlung processes with single and double parton-to-photon fragmentations.
Figure \ref{fig:DP} shows representative Feynman diagrams for DP diphoton plus dijet production.
For dijet scattering, the $gg\to gg$ process is shown, because it is dominant in the jet kinematic range studied in this analysis.

\begin{figure}[h]
\vspace*{0mm}
\hspace*{-5mm} \includegraphics[scale=0.42]{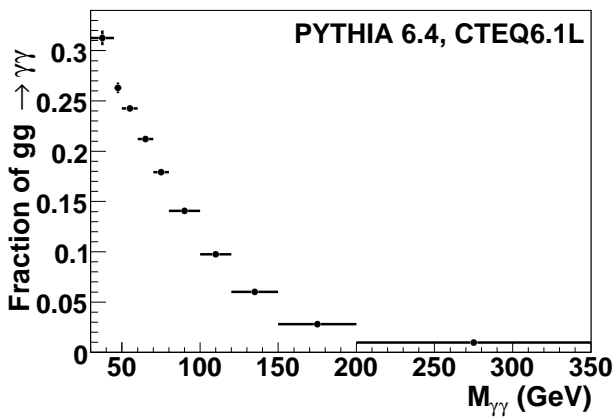}
\caption{
Fraction of the  $gg \to \gamma\gamma$ contribution to the total direct cross section comprising
the $q\bar{q} \to \gamma\gamma$ and $gg \to \gamma\gamma$ processes. $M_{\gamma \gamma}$ is the invariant mass of the diphoton.
}
\label{fig:qg_frac}
\vspace*{5mm}
\end{figure}

Figure~\ref{fig:qg_frac} shows the relative fraction of the $gg \to \gamma\gamma$ contribution to the total diphoton
cross section, which is a combination of $q\bar{q} \to \gamma\gamma$ and $gg \to \gamma\gamma$ processes.
For this analysis, which restricts the transverse momenta of each of the two leading jets to the range of 15--40 GeV and the transverse momenta of each of the two  leading photons to be above 15 GeV,
the $q \bar q$ scattering significantly dominates
the $gg$ process, with $q \bar q$ fraction of about 70\%--80\%.

 The outline of the paper is as follows. 
Section \ref{sec:Method} briefly describes the method for extracting \sigmaeff~
proposed in Ref.~\cite{CDF97}.
Section \ref{sec:ObjectID} introduces the D0 detector and data samples.
 Section \ref{sec:Models} describes
the signal and background models used in this measurement. 
 Section \ref{sec:Vars} discusses the discriminating
variable used to identify a data sample with an enhanced population of DP events.
The procedure for finding the fraction of DP events is given in Sec. \ref{sec:DPfraction}. 
 Section \ref{sec:DIfraction} contains a description of the analogous
procedure used to measure the fraction of events with double $p\bar{p}$ interactions.
A summary of the efficiencies required for the measurement is presented in Sec. \ref{sec:Eff}.
In Sec. \ref{sec:Sigma_eff}, we calculate the effective cross section,
$\sigma_{\rm eff}$, for the diphoton+dijet final state. 
The conclusions and outlook are presented in
Sec. \ref{sec:summary}. 

\section{Technique for extracting $\sigma_{\rm eff}$ from data}
\label{sec:Method}

The technique for extracting $\sigma_{\rm eff}$ has been used
in a number of earlier measurements~\cite{CDF97,D02010,D02014}.
To avoid using theoretical predictions for the SP diphoton and dijet cross sections,
the technique is based on a comparison of the number of \ggjj events produced in DP interactions
in single \ppbar collisions to the number of \ggjj events produced in two separate \ppbar collisions.
In the latter class of events, referred to as double interaction (DI) events, 
two hard parton interactions occur in exactly two separate \ppbar collisions within the same beam crossing. 

The single~\cite{CDF_hSD,D0_hSD} and double~\cite{D0_hDD} diffractive processes
contribute approximately 1\% to the total dijet production cross section with jet $p_T\gtrsim 15$ GeV.
Therefore, the diphoton and dijet events are produced mainly as a result of  inelastic 
nondiffractive (hard) $p\bar{p}$ interactions. 
In a $p\bar{p}$
beam crossing with two inelastic nondiffractive collisions the probability for a DI event is
\begin{equation}
P_{\rm DI} = 2~\frac{\sigma^{\gamma \gamma}}{\sigma_{\rm hard}} \frac{\sigma^{jj}}{\sigma_{\rm hard}},
\end{equation} 
where ${\sigma^{\gamma \gamma}}/{\sigma_{\rm hard}}$ ~(${\sigma^{jj}}/{\sigma_{\rm hard}}$) is
the probability for producing a diphoton (dijet) event satisfying particular photon (jet) selection criteria
in two separate  hard processes and $\sigma_{\rm hard}$ is the cross section of the hard $p \bar p$ interactions.
The factor of 2 accounts for the fact that the two scatterings (producing diphoton and dijet events) 
can be ordered in two ways with respect to the two collision vertices. 
The number of DI events can be obtained
from $P_{\rm DI}$, after correcting for geometric and kinematic acceptance $A_{\rm DI}$, selection efficiency (including trigger efficiency) $\epsilon_{\rm DI}$,
and the two-vertex selection efficiency $\epsilon_{\rm 2vtx}$ 
and multiplying by  the number of beam crossings with exactly two hard collisions $ N_{c}(2)$:
\begin{eqnarray}    
N_{\rm DI} = 
~2 \frac{\sigma^{\gamma \gamma}}{\sigma_{\rm hard}} \frac{\sigma^{jj}}{\sigma_{\rm hard}}
~N_{c}(2) ~A_{\rm DI} ~\epsilon_{\rm DI} ~\epsilon_{\rm 2vtx}.
\label{eq:ndi}
\end{eqnarray}

Similarly to $P_{\rm DI}$, the probability for DP events, $P_{\rm DP}$, in a beam crossing with one hard
collision, using Eq.~(\ref{eq:sigmaeff_main}), is,
\begin{equation}
P_{\rm DP} = \frac{\sigma^{\rm DP}}{\sigma_{\rm hard}} = 
\frac{\sigma^{\gamma \gamma}}{\sigma_{\rm eff}} \frac{\sigma^{jj}}{\sigma_{\rm hard}}.
\end{equation} 
The parton scatterings in the DP events are assumed to be uncorrelated ~\cite{Landsh,Goebel,TH1,TH1b,TH2,TH2b,TH3,Threl,Threlb}.
The number of DP events, $N_{\rm DP}$, can be expressed as $P_{\rm DP}$
corrected for the acceptance $A_{\rm DP}$, selection efficiency (including trigger efficiency) $\epsilon_{\rm DP}$,
and the single vertex selection efficiency $\epsilon_{\rm 1vtx}$, 
multiplied by the number of beam crossings with exactly one hard collision $ N_{c}(1)$:
\begin{eqnarray}    
N_{\rm DP} =
~\frac{\sigma^{\gamma \gamma}}{\sigma_{\rm eff}} \frac{\sigma^{jj}}{\sigma_{\rm hard}}
~N_{c}(1) ~A_{\rm DP} ~\epsilon_{\rm DP} ~\epsilon_{\rm 1vtx}.
\label{eq:ndp}
\end{eqnarray}   

Taking the ratio $N_{\rm DI}/N_{\rm DP}$ allows one to obtain an
expression for $\sigma_{\rm eff}$,

\begin{eqnarray}   
\sigma_{\rm eff} = 
\frac{N_{\rm DI}}{N_{\rm DP}}
\frac{A_{\rm DP}}{A_{\rm DI}}\frac{\epsilon_{\rm DP}}{\epsilon_{\rm DI}} \frac{\epsilon_{\rm 1vtx}}{\epsilon_{\rm 2vtx}} R_c \thinspace \sigma_{\rm hard}, 
\label{eq:sig_eff}
\end{eqnarray}
where $R_c =  {N_{c}(1) }/{2N_{c}(2)}$.

It is worth noting that (a) the $\sigma^{\gamma \gamma}$ and $\sigma^{jj}$ cross sections cancel in this ratio
and (b) the efficiencies and acceptances for DP and DI events enter only as ratios (i.e. all common
uncertainties are reduced as well). To calculate these efficiencies, acceptances, and their ratios,  
we use the data based models which are described in Sec.~\ref{sec:SignalModels}. 

The numbers of DI (DP) events $N_{\rm DI}$ ($N_{\rm DP}$)
can be determined from the number of two- (one-)vertex \ggjj events $N_{\rm 2vtx}$ ($N_{\rm 1vtx}$)
as 
$N_{\rm DI} = f_{\rm DI} P^{\gamma\gamma}_{\rm DI} N_{\rm 2vtx}$ ($N_{\rm DP} = f_{\rm DP} P^{\gamma\gamma}_{\rm DP} N_{\rm 1vtx}$),
where \fdi (\fdp) and $P^{\gamma\gamma}_{\rm DI}$ ($P^{\gamma\gamma}_{\rm DP}$) are the fraction of DI (DP) events and diphoton purity in the  two- (one-)vertex data set, respectively.
The fraction \fdp is estimated from the data set with one \ppbar collision using a fraction ratio
method, while \fdi can be obtained from data events with two \ppbar collisions 
using a jet-track algorithm. 
The complete description of the techniques used for \fdp and \fdi estimates are described in 
Secs.~\ref{sec:DPfraction} and \ref{sec:DIfraction}, and the diphoton sample purity is discussed in Sec.~\ref{sec:sigfrac}.

The main background for the DP events is due to contributions from
the SP scattering processes,
$q\bar{q} \to \gamma\gamma gg$, and $gg \to \gamma\gamma gg$.
These processes mainly result from gluon radiation in the initial or the final state
and can also result from photon fragmentation events.

\section{D0 detector and data samples}
\label{sec:ObjectID}

The D0 detector is described in detail in Refs.~\cite{d0det,l1cal,l0}. Photon candidates are
identified as isolated clusters of 
energy depositions in one of three uranium and liquid argon sampling calorimeters. 
The central calorimeter  covers the pseudorapidity range $|\eta_{\rm det}|<1.1$,
and the two end calorimeters extend the coverage up to $|\eta_{\rm det}| \approx 4.2$.
In addition, the plastic scintillator intercryostat detector covers
the region $1.1<|\eta_{\rm det}|<1.4$.
The electromagnetic (EM) section of the calorimeter is segmented longitudinally into 
four layers and transversely into cells in pseudorapidity and azimuthal angle 
$\Delta\eta_{\rm det}\times\Delta\phi_{\rm det}=0.1 \times 0.1$ 
($0.05 \times 0.05$ in the third layer of the EM calorimeter).
The hadronic portion of the calorimeter is located behind the EM section.
The calorimeter surrounds 
a tracking system consisting of a silicon microstrip tracking  detector and
scintillating fiber tracker, both located within a 1.9~T solenoidal magnetic field.
The solenoid magnet is surrounded by the central preshower (CPS)
detector located immediately before the calorimeter. The CPS consists of
approximately one radiation length of lead absorber at normal incidence surrounded by three layers of scintillating strips.
The luminosity of colliding beams
is measured using plastic scintillator arrays installed in
front of the two end calorimeter cryostats~\cite{andeen}.

The current measurement is based on 8.7~fb$^{-1}$ of data collected using $p\bar{p}$ collisions at
$\sqrt{s}=1.96$ TeV
after the D0 detector upgrade in 2006~\cite{l0}, 
while the previous measurements~\cite{D02010, D02011} were made using the data collected before this upgrade. 
The events used in this analysis pass the triggers designed to identify
high-\pt clusters in the EM calorimeter with loose
shower shape requirements for photons.
These triggers have $\approx 90\%$ efficiency for a photon transverse momentum $\Ptg \approx 16$~GeV and are $100\%$ efficient
for $\Ptg \gt 35$~GeV.

To select photon candidates in our data samples,
we use the following criteria~\cite{gamjet_tgj,EMID_NIM}:
EM objects are reconstructed using a simple cone algorithm with
a cone size of $\Delta{\cal R}=\sqrt{(\Delta\eta)^2 + (\Delta\phi)^2}=0.2$.
Regions with poor photon identification and degraded $\Ptg$ resolution
at the boundaries between calorimeter modules and between the central and end cap
calorimeters are excluded from the analysis.
Each photon candidate is required to deposit more than 96\% of the detected energy
in the EM section of the calorimeter 
and to be isolated in the angular region between
$\Delta{\cal R}=0.2$ and $\Delta{\cal R}=0.4$ around the 
center of the cluster:
$(E^{\rm iso}_{\rm tot}-E^{\rm iso}_{\rm core})/E^{\rm iso}_{\rm core} < 0.07$, where $E^{\rm iso}_{\rm tot}$ 
is the total (EM+hadronic) tower energy in the ($\eta,\phi$) cone of radius $\Delta{\cal R}=0.4$
and $E^{\rm iso}_{\rm core}$ is EM energy within a radius of $\Delta{\cal R}=0.2$.
Candidate EM clusters that match to a reconstructed track are excluded from the analysis. 
We also require the energy-weighted EM cluster width in the finely segmented third EM layer
to be consistent with that expected for a photon-initiated electromagnetic shower.
In addition to the calorimeter isolation cut, we also apply a track isolation cut,
requiring the scalar sum of the track transverse
momenta in an annulus $0.05 \leq \Delta{\cal R} \leq 0.4$ to be less than 1.5~GeV.
To further suppress the jet background, the photons are selected to satisfy the same requirement on a 
neural network (NN) discriminant as in Ref.~\cite{diphoton}.

Jets are reconstructed using 
an iterative midpoint cone algorithm~\cite{JetAlgo}
with a cone size of $0.7$. Jets must satisfy
quality criteria that suppress background from leptons, photons, and
detector noise effects.
Jet transverse momenta are corrected to the particle level~\cite{JES_NIM}.

Two photons must be separated from each other by $\Delta{\cal{R}}>0.4$ and 
from each jet by $\Delta{\cal{R}}>0.9$.
Jets must be separated from each other by $\Delta{\cal{R}}>1.4$.
Each event must contain at least two photons in the pseudorapidity
region $|\eta^{\gamma}|<1.0$ and at least two jets with $|\eta^{\rm jet}|<3.5$.
The photon with the highest \pt is named the ``leading photon,'' or first photon,
and the photon with the second highest \pt is denoted as 
the second photon.
Similar terminology is applied to the jets.
Events are selected with the leading photon transverse momentum 
$p^{\gamma}_{T}>16$~GeV, the second photon $p^{\gamma}_{T}>15$~GeV, 
and jets satisfying $15<p_T^{\rm jet}<40$~GeV.
The upper requirement on the \pt of the jets increases the
fraction of DP events in the sample~\cite{D02010}.
The numbers of events with exactly one
identified \ppbar collision (1VTX), exactly two identified \ppbar collisions (2VTX), and their ratio are shown in Table~\ref{tab:nev}.
The \ppbar collision vertices are reconstructed using a track-based algorithm and are sorted according to their tracking activity. 
The vertices are required to be within  $|z| \lt 60$~cm from the geometric center of the detector (the detector luminous region rms is $\sim 20$ cm) and have $N_{trk} \ge 3$ tracks.
The vertex at the top of the list (PV0) and the second-best (PV1) vertex have the highest and the second-highest tracking multiplicities, respectively.

\begin{table}[htpb]
\begin{center}
\caption{The number of selected \ggjj events with a single \ppbar collision ($N_{\rm 1vtx}$), two
\ppbar collisions ($N_{\rm 2vtx}$), and their ratio.}
\label{tab:nev}
\vskip 2mm
\begin{tabular}{ccc} \hline\hline
      $N_{\rm 1vtx}$ & $N_{\rm 2vtx}$ & $N_{\rm 2vtx}/N_{\rm 1vtx}$   \\\hline
      401  &   442   &   1.102    \\\hline\hline
\end{tabular}
\end{center}
\end{table}
\section{Data, signal, and background event models}
\label{sec:Models}

This section presents an overview of the DP and DI models built using
data and Monte-Carlo (MC) samples to estimate the number of DP and DI events in data,~\ndp and \ndi.
These models are also used to estimate the selection efficiencies and 
geometric and kinematic acceptances for DP and DI events.

\subsection{Signal models}
\label{sec:SignalModels}

Because \sigmaeff~depends on DP and DI events as shown in Eq.~(\ref{eq:sig_eff}),
both classes of events are considered signal events:
\vspace{-3mm}
\begin{enumerate}[(i)]
\itemsep-0.3em 
\item DP data event model (\mixdp): The DP event model 
is constructed by combining photons and jets from two events drawn from two samples: (a) an inclusive 
data sample of \gamgam events and (b)
a sample of 
inelastic nondiffractive events selected with a minimum bias trigger (a trigger that only requires hits in the luminosity detectors) and
a requirement of at least one reconstructed jet (``MB'' sample)~\cite{D02010,JES_NIM}. 
Both input samples contain events with exactly one reconstructed \ppbar collision vertex.
The resulting mixed event is required to satisfy the same selection criteria
as applied to \ggjj data events with a single \ppbar collision.
By construction, the \mixdp sample
provides independent parton scatterings with \gamgam and dijet final states.
Because the \gamgam process in a DP event is dominated by small parton momentum fractions ($x$),
the $x$ values in the dijet production process remaining after
the first parton interaction occurs are expected to be generally unaffected; 
i.e., the two interactions have negligible correlation in momentum space.
We have verified that the effect of adding the diphoton and dijet components in MIXDP with different vertex positions is negligible, 
since the MIXDP model is only used for modelling the transverse discriminating variable introduced below in Sec.~\ref{sec:Vars}.
Two possible event configurations with the \ggjj final state in a single $p\bar{p}$ collision
are shown in Fig.~\ref{fig:dp_mix}. 
\begin{figure}[h]
\includegraphics[scale=0.33]{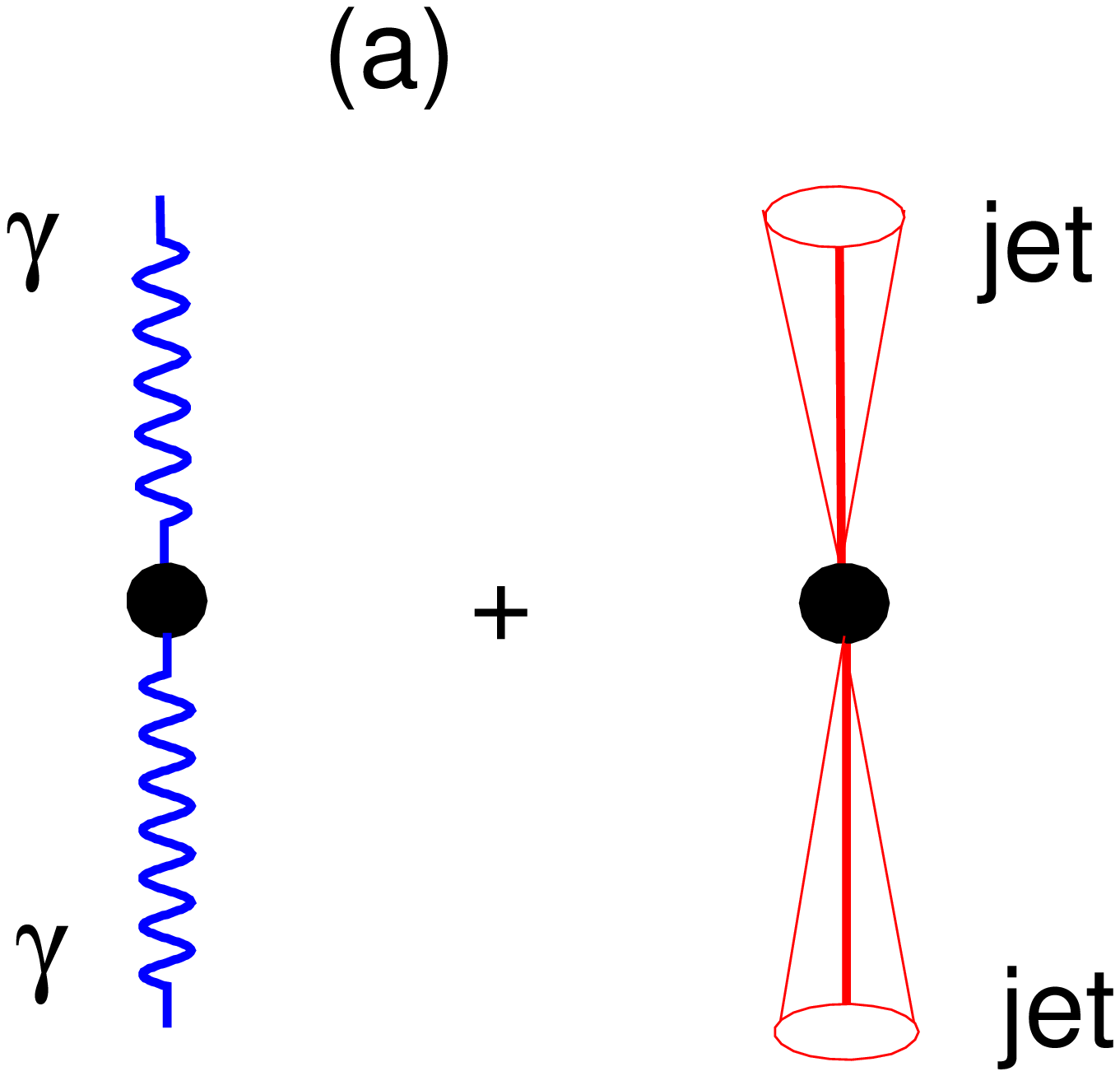}
\includegraphics[scale=0.33]{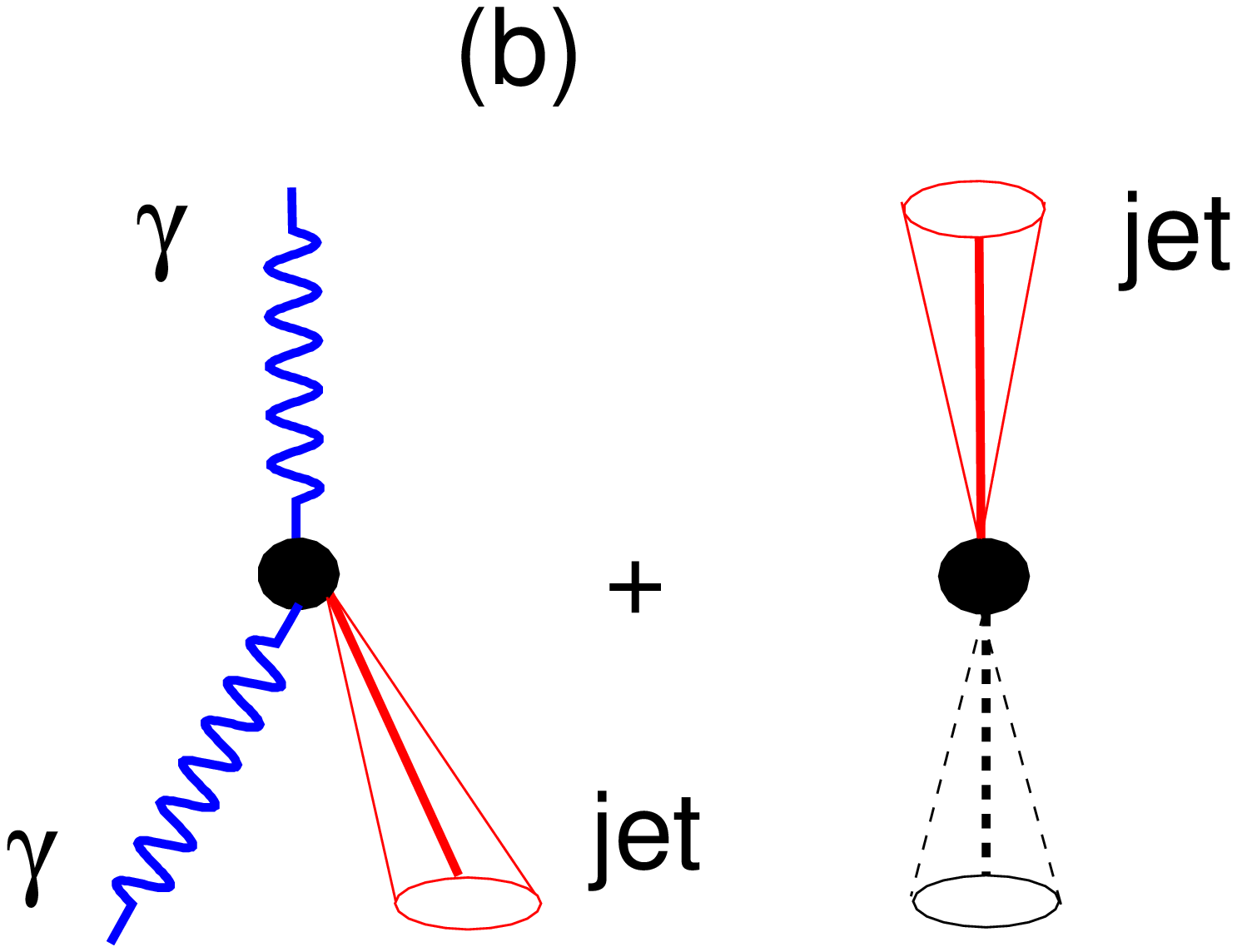}
\caption{ Diagrams of \ggjj final state in the events with a single $p\bar{p}$ collision.
{\it (a)} DP scattering with diphoton production overlaid with dijet production; 
{\it (b)} DP scattering with diphoton $+1$ jet production
overlaid with dijet production, in which one of the two jets is lost (dotted line).
They can also be used as an illustration of the two DI events if one assumes that the processes shown
come from two distinct $p \bar p$ collisions.
}
\label{fig:dp_mix}
\end{figure}
\item DI data event model (\mixdi): The \ggjj DI signal event model is built from 
an overlay of \gamgam and MB events with $\geq$1 
selected jets. This sample is prepared similarly to the \mixdp sample but with the 
requirement of exactly two reconstructed \ppbar collision vertices in both data samples instead of one such vertex in the samples used for \mixdp.
Thus, the second \ppbar collision contains only soft underlying energy
that can contribute energy to a jet cone, or a photon isolation cone.
In addition, in the case of jets in the MB component of the \mixdi mixture,
if there is more than one jet, both jets are required to originate 
from the same vertex, using jet-track information, 
as discussed in Appendix B of Ref.~\cite{D02010}.
The resulting \ggjj events undergo the same selection as
applied to the data sample with two \ppbar collision vertices.
\item DP and DI MC models (\mcdp and \mcdi): To create signal MC models for DP and DI events, 
we use an overlay of MC \gamgam and dijet events.
These events are generated with the
{\sc sherpa}~\cite{SHERPA} and {\sc pythia}~\cite{PYT} event generators, respectively, and are processed
by a {\sc geant}-based~\cite{GEANT} simulation of the D0 detector
response.  To accurately model the effects of multiple
\ppbar interactions and detector noise, data events from random
\ppbar crossings are overlaid on the MC events using data from the same data taking
period as considered in the analysis.
These MC events are then processed using the same reconstruction software as for data.
We also apply additional smearing to the reconstructed photon and jet \pt
so that the measurement resolutions in MC match those in the data.
These MC events are used to create single- and two-vertex samples. 

Using the \gamgam and dijet MC samples, we create 
\ggjj DP and DI MC models,
similarly to those constructed for \mixdp and \mixdi data samples, i.e., with only one and only two reconstructed primary interaction vertexes, respectively,
by examining information for jets and the photon
at both the reconstructed and particle level.
These samples are used to calculate selection efficiencies and acceptances for DP and DI events. 
As a cross check, we have compared the
\pt and $\eta$ distributions of the jets and photons at the reconstructed
level in these models with those
in the \mixdp and \mixdi data samples. Small discrepancies have been
resolved by reweighting these MC spectra and creating models denoted as datalike \mcdp and \mcdi.

\end{enumerate}

\subsection{Background model}

To extract the DP signal from the data, 
we need to subtract $\gamma \gamma +$dijet SP background.
\vspace{-3mm}
\begin{enumerate}[(i)]
\item SP one-vertex event model (\sponevtx): A background to the DP events arises predominantly from $\gamma \gamma$ production 
with two jets, resulting in a \ggjj final state
in a single \ppbar collision event. 
To model this background, we consider a sample of MC \ggjj events
generated with {\sc pythia} and {\sc sherpa} with multiple parton interaction  modeling turned off.
The \sponevtx sample contains the final state with two photons and two additional
 jets with the same selection criteria as applied to the data
sample with a single \ppbar collision vertex. Other small backgrounds 
 are included in the event generators.
The {\sc sherpa} SP model is taken as the default.
\end{enumerate}

\section{Discriminating variable}
\label{sec:Vars}

\begin{figure}[htp!]
\hspace*{2mm} \includegraphics[scale=0.35]{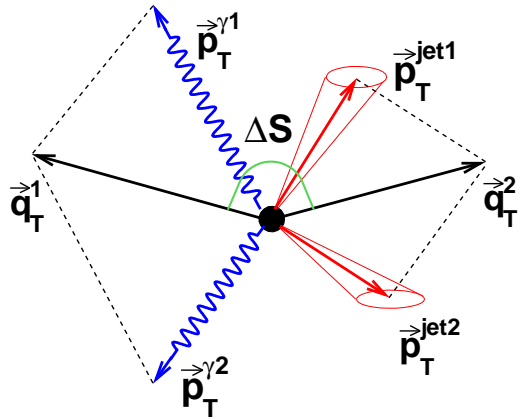}
\caption{
A diagram illustrating the orientation of photon and jet transverse momenta vectors in \ggjj events.
Vectors $q_T^{~1}$ and $q_T^{~2}$ are the $p_T$ imbalance vectors of diphoton and dijet pairs, respectively.  }
\label{fig:deltaPHI}
\end{figure}

A DP event contains two independent $2 \rightarrow 2$ 
parton-parton scatterings within the same \ppbar collision. 
The same final state can be produced by the SP $2 \rightarrow 4$ process, resulting in
\gamgam and two bremsstrahlung jets with substantially different kinematic distributions.
Discrimination between these processes is obtained by exploiting the
azimuthal angle between the \pt imbalance vectors of photon and jet pairs 
in \ggjj events,
\begin{eqnarray}
\Delta S \equiv \Delta\phi\left(\vec{q}_{\rm T}^{~1},~\vec{q}_{\rm T}^{~2}\right),
\label{eq:deltaPHI}
\end{eqnarray}
where $\vec{q}_{\rm T}^{~1} = \vec{p}_{\rm T}^{~\gamma_1} + \vec{p}_{\rm T}^{~\gamma_2}$
and $\vec{q}_{\rm T}^{~2} = \vec{p}_{\rm T}^{\rm ~jet_1} + \vec{p}_{\rm T}^{\rm ~jet_2}$.
Figure~\ref{fig:deltaPHI} illustrates the orientation of photons and 
jets transverse momentum vectors in \ggjj events, as well as the imbalance vectors
$\vec{q}_{\rm T}^{~1}$ and $\vec{q}_{\rm T}^{~2}$.

For DP events in which the photons come from one parton-parton scattering and 
the two jets come from another
parton-parton scattering, the \dS angle is isotropically distributed.
However, the DP events 
with an additional bremsstrahlung jet in the first parton-parton scattering shown in Fig.~\ref{fig:dp_mix}(b)
tend to populate the region toward \dS$=\pi$ due to momentum conservation.
The bremsstrahlung processes also cause \dS to peak strongly near $\pi$
in SP, but
detector resolution effects and gluon radiation in parton showers produce a tail
extending to smaller angles.
\section{Fractions of DP and DI events}
\label{sec:Frac}

\subsection{Fractions of DP events}
\label{sec:DPfraction}

In order to calculate \sigmaeff, one needs to measure the number of DP events (\ndp) 
which enters Eq.~(\ref{eq:sig_eff}), as the product of the fraction of DP events (\fdp) in the 1VTX data sample,
the size of the 1VTX sample, and its diphoton purity.
The fraction is estimated in the \ggjj 1VTX data sample using
the \mixdp and the \sponevtx models described in Sec.~\ref{sec:Models}.

The observed number of data events, $N^n_{\rm data}$, with $\Delta S$ less than a cut $\Delta S^n$ can be written as
\begin{equation}
    N^n_{\rm data} = f^n_{\rm DP} N^n_{\rm DP} + (1-f^n_{\rm DP}) N^n_{\rm SP}, \nonumber
\end{equation}
where the number of DP events normalized to the data sample is
$N^n_{\rm DP} = (N^{\rm tot}_{\rm data}/M^{\rm tot}_{\rm DP}) M^n_{\rm DP}$,  $N^{\rm tot}_{\rm data}$ and $M^{\rm tot}_{\rm DP}$
 are the total number of events in the data and \mixdp samples for all values of $\Delta S$, and $M^n_{\rm DP}$ is the number of \mixdp
 events below the cut $\Delta S^n$.  A similar construction is used to define $N^n_{\rm SP}$ using the \sponevtx sample. 
 We define the fractions $\epsilon^n_{\rm data} = N^n_{\rm data}/N^{\rm tot}_{\rm data}$, $\epsilon^n_{\rm DP} = N^n_{\rm DP}/N^{\rm tot}_{\rm DP}$,
 and $\epsilon^n_{\rm SP}=N^n_{\rm SP}/N^{\rm tot}_{\rm SP}$ and use the fact that $N^{\rm tot}_{\rm DP} = N^{\rm tot}_{\rm SP} = N^{\rm tot}_{\rm data}$ to obtain
\begin{equation}
    \epsilon^n_{\rm data} = f^n_{\rm DP} \epsilon^n_{\rm DP} + (1 -f^n_{\rm DP}) \epsilon^n_{\rm SP}, \nonumber
\end{equation}
which yields 
 \begin{equation}
    f^n_{\rm DP} = \frac{\epsilon^n_{\rm data} - \epsilon^n_{\rm SP}}{\epsilon^n_{\rm DP}-\epsilon^n_{\rm SP}}.
\label{eq:fdp_eff}
\end{equation}

Due to the definitions of the fractions $\epsilon^n$, this expression for $f^n_{\rm DP}$ depends upon the numbers of events in the data,
 DP, and SP distributions both below and above the cut, $\Delta S^n$.  
To estimate the uncertainties in the shapes of the $M_{\rm DP}$ and $M_{\rm SP}$
 distributions of \mixdp and \sponevtx events, respectively, as a function of $\Delta S$, we compute $f^n_{\rm DP}$ for seven different values of the cut value $\Delta S^n$, 
and average the results, taking into account the correlations in the numbers of events in the different samples. 
We also estimate the uncertainty due to model dependence of the \sponevtx sample as in the appendix of Ref.~\cite{D02014} by reweighting the models to data, based on the kinematic distribution $\Delta\phi(\gamma,\gamma)$ and the jet $p_T$ spectra. The differences between estimates made with the original and the modified models are included in the systematic uncertainty.  
The background due to DP photon-3jet events is corrected for using the diphoton purity estimate; see Sec.~\ref{sec:sigfrac}. 
Using an inclusive $\gamma + $ jet sample~\cite{gamjet_tgj}, we estimate the fraction of DP $\gamma+$ jet events to be less than 2.0\%.
  We do not correct for this effect and include the entire estimate of the contamination as a systematic uncertainty.
Finally, we get
\begin{equation}
    f^{\rm avg}_{\rm DP} = 0.213 \pm 0.061{\rm(stat)} \pm 0.028{\rm(syst)}.
\label{eq:fdp_ave}
\end{equation}

As a cross check, the fraction \fdp is found using a maximum likelihood fit~\cite{HMCMLL}
of the \dS distribution of the data to signal and background templates
that are taken to be the shapes of $M_{\rm DP}$ and $M_{\rm SP}$, respectively.
Signal and background models are described in Sec.~\ref{sec:Models}
and undergo all the selection criteria applied to the data sample.
From the fit
we find a $f_{\rm DP}$ value of $0.18 \pm 0.11$, which agrees with the value estimated by the average fraction
method within uncertainty.
\begin{figure}[htb!]
\vspace*{-3mm}
\hspace*{4mm} \includegraphics[scale=0.4]{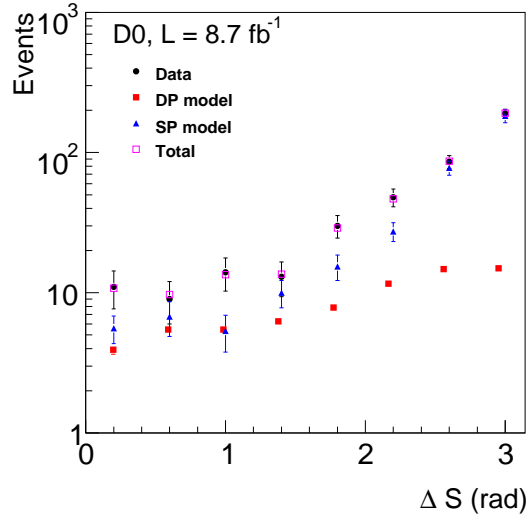}
~\\[-3mm]
\caption{ The fit of the 1VTX data \dS distribution with SP and DP templates to extract the DP fraction. 
The black points correspond to data, red boxes to the DP signal \mixdp model
normalized to the \fdp fraction obtained from the fit, and the blue triangles are the SP background template
(\sponevtx)  normalized to its fraction ($1 - $\fdp).
The pink open boxes correspond to the sum of the signal and background (total).
}
\label{fig:dpfit_init}
\end{figure}
The result of the fit is shown in Fig.~\ref{fig:dpfit_init}.

\subsection{Fractions of DI events}
\label{sec:DIfraction}

Double interaction events in the 2VTX sample arising from different $p\bar p$ interactions
 within the same bunch crossing include those events in which the $\gamma\gamma$ and dijets are associated with different vertices
 and those in which the two jets are associated with different vertices irrespective of the photons' vertex associations.
   Backgrounds to the DI events in the two-vertex sample come from those events in which the two photons and the two jets
 are associated with the same vertex (and there is an additional MB vertex containing neither a 
$\gamma$ nor jet).  The DI fraction, $f_{\rm DI}$, is defined as the ratio of the number of DI events to the sum of the DI and background events.

The vertex association for jets is based on the $p_T$-weighted average, $<$$z_{\rm vtx}$$>$, of the $z$ positions (points of the closest approach to $z$ axis) of all tracks associated with the jet
  and the charged particle fraction (CPF) discriminant that measures the fraction of the total charged particle $p_T$ in each jet $i$ that is associated with vertex $j$,
\begin{eqnarray}
{\rm CPF(jet}_i,{\rm vtx}_j) = \frac{\sum_k p_T({\rm trk}_k^{{\rm jet}_i},{\rm vtx}_j)}{\sum_n\sum_l p_T({\rm trk}_l^{{\rm jet}_i},{\rm vtx}_n)},
\end{eqnarray}
where the sum is taken over tracks within the jet cone in the numerator and also over all vertices in the denominator.
For the calculation of $f_{\rm DI}$, we require each jet to contain at least two tracks and to satisfy CPF $> 0.65$ for one of the two vertices.   
Using a sample of $\gamma +$ jet events with exactly one observed vertex, we find the resolution in the $p_T$-weighted jet $z$ position to be $\sigma^{\rm jet}_z = 1.2$ cm.
  We require a valid jet to point to one of the vertices within $3 \sigma^{\rm jet}_z$.

The $z$-resolution of photons using only the information from the EM calorimeter is too coarse to be of use in making a vertex association.
  However, for those photons in which there is a good three-dimensional cluster seen in the CPS,
 the combined EM calorimeter and CPS position information provides a photon pointing resolution of $\sigma^\gamma_z = 3$ cm.
  We require a CPS tagged photon to point to one of the vertices within $3 \sigma^\gamma_z$.

The fraction of events in the total DI sample of 442 events (cf.~Table~\ref{tab:nev}) in which the two jets are associated
 with different vertices is 14.6\%.   In this estimate, no requirement on the photon vertex assignments is made. 
 Using an inclusive $\gamma + $ jet sample~\cite{gamjet_tgj}, we estimate the fraction of non-DI events in which a $\gamma +$
 jet is associated with each of the different vertices to be less than 0.5\%.

About one-quarter of all two-vertex events have CPS pointing information for both photons.
  Using this sample, we estimate that 4.7\% of the two-vertex events are DI events in which the diphotons
 are associated with one vertex and the dijet systems are associated with the other.
  Due to the small sample statistics and relatively large $\sigma^\gamma_z$, we assign a 50\% uncertainty on this component of $f_{\rm DI}$.
  Taking the two categories of DI events together, we find $f_{\rm DI} = 0.193 \pm 0.021 ({\rm stat}) \pm 0.028 ({\rm syst})$.

The DI fraction could depend on the distance in $z$ between the two vertices.
To study this effect, the distance between the two vertices is varied
up to 7$\sigma_z^{\rm jet}$, and the DI fraction is extracted with the requirement above.
Table~\ref{tab:dZjet} shows \fdi with respect to the distance between two
vertices, $\Delta z(PV0,PV1)$.
The difference between the default \fdi value and \fdi found when the distance 
between the two vertices is greater than $7\sigma_z^{\rm jet}$ is added to the systematic uncertainty.
The default choice corresponds to no restriction on $\Delta z(PV0,PV1)$.
\begin{table}[htpb!]
	\centering
	\small
	\caption{DI event fraction with respect to $\Delta z (PV0, PV1)$.}
	\label{tab:dZjet}
	\begin{tabular}{cc} \hline\hline
	$\Delta z (PV0, PV1)$ ~&~ \fdi \\\hline
	Default ~&~ $0.193 \pm 0.021({\rm stat}) \pm 0.028({\rm syst}) $ \\ 
	$\gt 3\sigma_z^{\rm jet}$ ~&~ $0.195 \pm 0.021({\rm stat}) \pm 0.028({\rm syst}) $ \\ 
	$\gt 5\sigma_z^{\rm jet}$ ~&~ $0.200 \pm 0.022({\rm stat}) \pm 0.028({\rm syst}) $ \\ 
	$\gt 7\sigma_z^{\rm jet}$ ~&~ $0.203 \pm 0.023({\rm stat}) \pm 0.028({\rm syst}) $ \\ \hline\hline
	\end{tabular}
\end{table}  
Finally, the DI fraction extracted is:
\begin{equation}
\label{eq:fdi}
f_{\rm DI} = 0.193 \pm 0.021~({\rm stat}) \pm 0.030~({\rm syst})
\end{equation}

\section{DP and DI efficiencies, $\bm R_{\bm c}$ and $\bm \sigma_{\bf hard}$}
\label{sec:Eff}

\subsection{Ratio of photon purity in DP and DI events}
\label{sec:sigfrac}

As mentioned in Sec.~\ref{sec:Method}, the numbers of events $N_{\rm DI}$ and $N_{\rm DP}$
in Eq.~(\ref{eq:sig_eff}) depend on the purity of the diphoton sample. 
There are two major sources of background events to direct diphoton production: (i) Drell-Yan
(DY) events with both electrons misidentified as photons due to tracking inefficiency
and (ii) \gpj and dijet events with jet(s) misidentified as photon(s)~\cite{diphoton}.
The $W + {\rm jet}/\gamma$ background
with $W \to e\nu$ decay has been estimated from MC and is found to be negligible.
The number of data events that satisfy the photon selection criteria can be written as the sum of true diphoton events, 
DY events and \gpj or dijet events that fake the two photon signature.

We use $Z/\gamma^{*} \to ee$ {\sc pythia+alpgen} MC samples to estimate the DY contribution. 
The next-to-next-to-leading-order $p\bar{p} \to Z/ \gamma^{*} \to ee$ cross section~\cite{zdy} 
is used for the absolute normalization and the generator
level $Z/\gamma^{*}$ boson \pt has been reweighted to the measured data distribution. The expected number of
events from the DY process is $2.19 (0.5\%)$ and $2.41 (0.5\%)$ in case of 1VTX and 2VTX events, respectively.
The numbers in parentheses correspond to the percentage of the DY contribution to the data sample.

To estimate the fraction of diphoton events, we use variables 
sensitive to the internal structure of the electromagnetic shower. 
The outputs of the photon NN~\cite{diphoton} for the photons in the central calorimeter, 
trained on MC samples with direct photons and dijets,
have been chosen as a discriminant between signal and background events.
Since the signal events cannot be identified on an event by event basis,
their fraction (purity) 
$P^{\gamma\gamma}$, defined as the ratio of the number of two photon events
to the total number of candidate events satisfying the selection criteria, is determined statistically.

The two-dimensional distribution of NN outputs of the two photon candidates in data
after subtracting the DY contribution
 is fitted using two-dimensional NN output templates of signal photons from the {\sc sherpa} and {\sc pythia} MC
 and templates of jets from {\sc pythia} MC jet samples,
where special requirements are applied at the generator level to enrich the sample with jets having an electromagnetic shower shape similar to that of the photon~\cite{diphoton}.
The fit uses the same maximum likelihood method \cite{HMCMLL} as for the cross check fit for \fdp; see Sec.~\ref{sec:DPfraction}.
The results of the diphoton purities in DP and DI events and their ratio are presented in Table~\ref{tab:pur_fit}.

\begin{table}[htpb]
\begin{center}  
\caption{Diphoton event purity in DP and DI events and their ratio. The uncertainties are statistical.}
\label{tab:pur_fit}
\vskip 1mm
\begin{tabular}{ccc} \hline \hline
 Sample     & \sherpa & \pythia \\\hline 
  $P^{\gamma\gamma}_{\rm DP}$         &  0.688$\pm$0.005  & 0.608$\pm$0.028    \\
  $P^{\gamma\gamma}_{\rm DI}$         &  0.689$\pm$0.025  & 0.623$\pm$0.029   \\ 
  $P^{\gamma\gamma}_{\rm DI}/P^{\gamma\gamma}_{\rm DP}$    &  1.002$\pm$0.039  & 1.025$\pm$0.067   \\\hline\hline
\end{tabular}
\end{center} 
\end{table} 

We identify an additional source of systematic uncertainty due to model dependence as half of the difference between 
the ratio of purities calculated using different signal models generated by \pythia and \sherpa.
It is estimated to be $1.2\%$.

Another source of systematic uncertainty is due to the fragmentation model used in \pythia
and caused by the uncertainty in the fragmentation functions $D_{\pi, \eta}(z)$. 
This uncertainty is estimated by varying the number of $\pi^0$
and $\eta$ mesons in the dijet sample by a factor of 2 and calculating the purity using
the modified templates. It is found to be equal to $3\%$.

\subsection{Ratio of geometric acceptance times efficiency in DP and DI events}

The acceptance ($A$) is calculated as a ratio of ${N^{\rm reco}_{i}}/{N^{\rm gen}_{i}}$,
where $N^{\rm reco}_{i}$ and $N^{\rm gen}_{i}$ are the numbers of simulated events
at the reconstruction and generator (true) level, respectively. 
It accounts for events lost during event reconstruction, for objects created by spurious hits, and the contribution from true objects outside the fiducial region but reconstructed 
inside the fiducial region and vice versa.

To estimate acceptances in one and two \ppbar collision samples,
we use the signal \mcdp and \mcdi samples described in Sec.~\ref{sec:Models}.
These samples mix diphoton events generated by \sherpa and dijet events
generated by \pythia.
The acceptance is calculated using the following photon and jet selection criteria:
\vspace{-3mm}
\renewcommand{\labelenumi}{(\arabic{enumi}) }
\begin{enumerate}
\itemsep-0.3em 
\item Generator level:
\vspace{-3mm}
\begin{enumerate}
\itemsep-0.5em 
\item
\ptgamone $> 16$ \GeV, \ptgamtwo $> 15$ \GeV, $|\eta|<1.0$; \\
\item
jets with $15 \lt p_{T}^{\rm jet}\le 40$ \GeV and $|\eta^{\rm jet}| \lt 3.5$;
\end{enumerate}
  \item Reconstruction level: 
\vspace{-3mm}
\begin{enumerate}
\itemsep-0.5em 
\item
\ptgamone $> 16$ \GeV, \ptgamtwo $> 15$ \GeV, $|\eta|\lt 1.0$, $|\eta^{\rm det}|\lt 1.0$; \\
photon candidates are required to be away from the calorimeter module boundaries in $\phi^{\rm det}$;
the fraction of the photon energy in the EM calorimeter is required to be greater than 0.9;
and the fraction of energy in the calorimeter isolation annulus $0.2 < \Delta R < 0.4$ around the photon 
is required to be 0.15 of that within the $\Delta R=0.2$ cone;\\
\item
jets with $15 \lt p_{T}^{\rm jet}\le 40$ \GeV, $|\eta^{\rm jet}| \lt 3.5$.
\end{enumerate}
\end{enumerate}
In Table~\ref{tab:acc_value}, we present the photon and jet acceptance
for 1VTX (MCDP) and 2VTX (MCDI) samples and their ratio.
\begin{table}[htpb]
\begin{center}
\caption{Geometric acceptances in DP and DI events and their ratio.}
\label{tab:acc_value}
\begin{tabular}{ccc} \hline\hline
        $A_{\rm DP}$ & $A_{\rm DI}$  & $A_{\rm DP}/A_{\rm DI}$   \\\hline
 0.429 $\pm$ 0.008    &  0.826 $\pm$ 0.019   &  0.521 $\pm$ 0.015 \\\hline\hline
\end{tabular}
\end{center}
\end{table}
The difference between 1VTX and 2VTX acceptances is mostly caused by 
different amounts of underlying energy falling inside the photon and jet cones,
resulting in different efficiencies for passing the photon and jet $p_T$ requirements.
The uncertainties due to the jet energy scale
(JES) and the model dependence of the individual acceptances largely cancel in the ratio.

\subsection{Ratio of photon efficiencies in DP and DI events}

The DP and DI events differ from each other by the number of $p\bar{p}$
collision vertices  (one vs.\ two), and therefore 
their selection efficiencies $\epsilon_{\rm DP}$ and $\epsilon_{\rm DI}$ may differ
due to different amounts of soft unclustered energy in the single and double
\ppbar collision events.
This could lead to different photon selection efficiencies because of
different distortions of the shower shape that this unclustered energy may introduce into the track and calorimeter isolation cones
around the photon. 

The efficiency for passing the photon selection criteria is estimated
using \ggjj \pythia and \sherpa MC events. 
The events are preselected with all jet cuts and loose photon identification cuts 
(as used in the acceptance calculation), and \onevtx and \twovtx samples are extracted from them.
The efficiency is calculated from the ratio of the number of events that pass the photon
selection criteria, weighted by the trigger efficiency to the number of events that
pass the preselection criteria.
In Table~\ref{tab:gameff_gj}, we present the photon efficiencies
for DP and DI events. Uncertainties are due to limited MC statistics.

\begin{table}[htpb]
\begin{center}
\caption{Photon efficiencies in single and double \ppbar collisions \ggjj \sherpa and {\pythia} MC samples. Uncertainties are due to limited MC statistics.}
\label{tab:gameff_gj}
\begin{tabular}{ccc} \hline
Sample     & \sherpa & \pythia  \\\hline\hline
$\epsilon_{\rm DP}$         &   0.477 $\pm$ 0.035    &  0.576 $\pm$ 0.010 \\
$\epsilon_{\rm DI}$         &   0.333 $\pm$ 0.021    &  0.419 $\pm$ 0.009 \\
$\epsilon_{\rm DP}/\epsilon_{\rm DI}$ & 1.434 $\pm$ 0.138    &  1.372 $\pm$ 0.039  \\\hline\hline
\end{tabular}
\end{center}
\end{table}

The difference in the efficiencies between {\sc pythia} and {\sc sherpa} is used
as an estimate of the systematic uncertainty due to model dependence.
The selection efficiencies for DP and DI events enter Eq.~(\ref{eq:sig_eff})
only as a ratio, substantially canceling correlated systematic uncertainties.
The {\sc pythia} ratio, which has a smaller statistical uncertainty, is used in the $\sigma_{\rm{eff}}$ calculation.

\subsection{Ratio of vertex efficiencies}

An efficiency, $\epsilon_{\rm 1vtx}$ ($\epsilon_{\rm 2vtx}$),  calculated for the DP (DI) candidate samples,
is mostly due to the single (double) vertex requirements, $|z| \lt 60$~cm and $N_{trk} \ge 3$.
The contribution of the vertex reconstruction efficiency to this quantity is partially absorbed into the acceptance calculation and very close to unity, as we discuss below.
 To calculate the efficiency for events with 1 \ppbar collision to pass the vertex requirement,
we use the \ggjj data with photon and jet selection criteria.
The efficiency for simultaneously satisfying the two-vertex requirement
is estimated separately for each jet-vertex assignment configuration, since the vertex
efficiency depends on the objects originating from the vertex.
For diphoton-dijet events originating from two separate vertices, we calculate $\epsilon_{\rm 2vtx}$
as a product of the efficiency to pass the vertex cuts in the
diphoton \twovtx data sample and 
the efficiency to pass the vertex cuts for dijets in the \twovtx MB sample.
Similarly, for events with two jets originating from two separate vertices, we calculate the $\epsilon_{\rm 2vtx}$ efficiency
as a product of the efficiency to pass the vertex cuts for the 
$\gamma\gamma + 1$ jet \twovtx data sample and    
the efficiency to pass the vertex cuts for jets in the \twovtx MB sample.
The final efficiency is a combination of the two, weighted by the event-type fraction.
Table~\ref{tab:vertex} presents the vertex efficiencies
for \onevtx and \twovtx samples and their ratio.

\begin{table}[htpb]
\begin{center}
\caption{Vertex efficiencies for \onevtx and \twovtx samples and their ratio.}
\label{tab:vertex}
\begin{tabular}{ccc} \hline\hline
        $\epsilon_{\rm 1vtx}$ & $\epsilon_{\rm 2vtx}$  & $\epsilon_{\rm 1vtx}/\epsilon_{\rm 2vtx}$   \\\hline
        0.944 $\pm$ 0.003   &  0.922 $\pm$ 0.003     &  1.021 $\pm$ 0.005      \\\hline\hline
\end{tabular}
\end{center}
\end{table}

We also estimate the probability to lose a hard interaction event because no primary vertex is reconstructed.
We find that the fraction of such events
in the MB event sample with jet $p_T \gt 15$~GeV is about 0.1\%
and about 0.2\% for the $\gamma \gamma + \geq 1$ jet events in data.
Due to the vertex reconstruction algorithm,
we may also have an additional reconstructed vertex that passes
the vertex requirement.
The rate at which this occurs is estimated using $\gamma \gamma + \geq 1$ jet events
and  $\gamma \gamma + \ge 2$ jets events simulated in MC without
the events from random $p \bar p$ bunch crossings overlaid (there should not be a second vertex in this case).
The probability to have a second vertex is around 0.05\%.
An analogous estimate for dijet events
(with the requirement of $\ge$ 1 and $\ge$ 2 jets) returns a probability of around 0.1\%.

\subsection{Correction of $N_{DI}$ for the track efficiency requirement}
\label{sec:ntrk2}

For the DI fraction calculation, we use the CPF algorithm, described in Sec.~\ref{sec:DIfraction}.
The method requires $\ge$ 2 tracks and returns the highest CPF.
The efficiency for the track requirement is calculated similarly to the vertex efficiency for each event-type
and then combined with the event type weights.
Finally, the estimated number of DI events, \ndi, is corrected for the $\epsilon_{Ntrk \ge 2}$ efficiency
which is found to be $\epsilon_{Ntrk \ge 2} = 0.725 \pm 0.004$.

\subsection{Calculating $\bm R_{\bm c}$, \boldmath{$\sigma_{\rm hard}$}, \boldmath{$N_{\rm 1coll}$} and \boldmath{$N_{\rm 2coll}$}}

We calculate the numbers of expected events with
one [$N_c(1)$] and two [$N_c(2)$] $p\bar{p}$ collisions 
resulting in hard interactions following the procedure of Ref.~\cite{D02010},
which uses the hard $p\bar{p}$ interaction cross section 
$\sigma_{\rm hard} = 44.76 \pm  2.89 ~{\rm mb}$.\ 
The values of $N_c(1)$ and $N_c(2)$ are obtained from 
a Poisson distribution parametrized with the average number of hard interactions in each
bin of the instantaneous luminosity, $L_{\rm inst}$, distribution, 
$\la n \ra = (L_{\rm inst} / f_{\rm cross} ) \sigma_{\rm hard}$,
where $f_{\rm cross}$ is the frequency of beam crossings
for the Tevatron~\cite{d0det}. 
Summing over all $L_{\rm inst}$ bins, weighted with their fractions,
we get $R_c = (1/2) (N_c(1) /N_c(2)) = 0.45$.
Due to higher instantaneous luminosities, this number is smaller by approximately a factor of 2 compared to that
for the data collected earlier as reported in Ref.~\cite{D02010}.
Since $R_c$ and $\sigma_{\rm hard}$ enter Eq.~(\ref{eq:sig_eff}) for $\sigma_{\rm eff}$
as a product, any increase of $\sigma_{\rm hard}$ leads to an increase of $\la n \ra$
and, as a consequence, to a decrease in $R_c$, and vice versa.
Although the measured 
value of $\sigma_{\rm hard}$ has a 6\% relative uncertainty, due to this partial cancellation of uncertainties, the product $R_c\sigma_{\rm hard}$
only has a 2.6\% uncertainty: $R_c\sigma_{\rm hard} = 18.92 \pm 0.49$ mb.

\section{Results}
\label{sec:Sigma_eff}

The uncertainty in the JES affects the ratio  $N_{\rm DI}/N_{\rm DP}$ in Eq.~(\ref{eq:sig_eff}).
We assess this uncertainty by raising and lowering JES by 1 GeV to give an uncertainty in $\sigma_{\rm eff}$ of 13.2\%.
We use Eq.~(\ref{eq:sig_eff}) to obtain \sigmaeff:
\begin{equation}
\label{eq:s_eff}
\sigma_{\rm eff} = 19.3 \pm 1.4 ({\rm stat}) \pm 7.8 ({\rm syst}) {\rm mb}.
\end{equation}
The main sources of systematic uncertainties are summarized
in Table~\ref{tab:syst}. The dominant sources are those due to \fdp, \fdi, and JES.
\begin{table*}[htpb]
	\caption{Systematic and statistical uncertainties (in \%).
        The contributions to the
	total systematic uncertainty come from uncertainties in the fraction of DP and DI in the one- and two-vertex events samples (\fdp and \fdi),
	the ratio of efficiency times acceptance (``EffRatio''),
	the ratio of photon fractions (``Purity''),
	JES, and the ratio of the number of events with single and double $p\bar{p}$ 
	hard collisions (``$R_{\rm c}\sigma_{\rm hard}$'').}
	\label{tab:syst}
	\begin{tabular}{ccccccccc} \hline\hline
  	 $f_{\rm DP}$  & $f_{\rm DI}$ & EffRatio & Purity & JES & $R_{\rm c}\sigma_{\rm hard}$ & SystTotal & StatTotal & Total \\\hline
           31.0 &  18.7 &   7.1 &   7.2 &   13.2 &   2.6 &    40.2 &     6.9 &    40.8 \\\hline\hline
  	\end{tabular}
\end{table*}

\begin{figure}[htp!]
\hspace*{2mm} \includegraphics[scale=0.45]{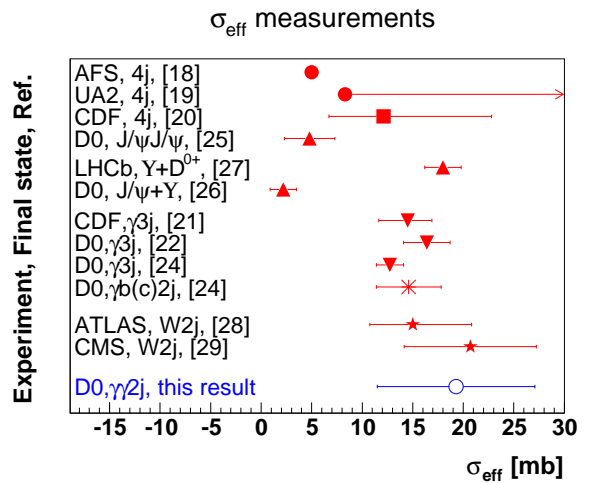}
\caption{
Existing  measurements of the effective cross section, $\sigma_{\rm eff}$, 
compared to the result presented here
(AFS: no uncertainty is reported; UA2: only a lower limit is provided).
Results of the measurements are grouped by the final state.
}
\label{fig:sigmaEff_world}
\end{figure}

Figure~\ref{fig:sigmaEff_world} shows all the measurements of \sigmaeff~
performed by various experiments up to the present time.
One can see that the \sigmaeff~ obtained by this measurement agrees with the recent D0 measurements \cite{D02010,D02014} 
and with those obtained by other experiments for processes dominated by $q \bar q$ and $qg$ initial states.

\section{Summary}
\label{sec:summary}
We have presented the first measurement 
of double parton scattering processes
in a single \ppbar collision
with \ggjj final states.
In the chosen kinematic region, \ptgamone $\gt$ 16 GeV, \ptgamtwo $\gt$ 15 GeV, $|\eta^{\gamma}|<1.0$, $|\eta^{\rm jets}|<3.5$, and
15 $< p^{\rm jets}_{\rm T} < $ 40 GeV, photon separation $\Delta{\cal{R}}>0.4$, photon-jet separation
$\Delta{\cal{R}}>0.9$, and jet-jet separation $\Delta{\cal{R}}>1.4$, we observe that $21.3 \pm 6.7$\% of 
events arises from double parton scattering.
The parameter \sigmaeff, which characterizes the size of the 
interaction region in a nucleon, is 
 found to be 
$\sigma_{\rm eff} = 19.3 \pm 1.4~({\rm stat}) \pm 7.8~({\rm syst})$ mb. 

~\\[5mm]
\centerline{\bf Acknowledgements}
~\\[1mm]

%

We thank the staffs at Fermilab and collaborating institutions,
and acknowledge support from the
Department of Energy and National Science Foundation (United States of America);
Alternative Energies and Atomic Energy Commission and
National Center for Scientific Research/National Institute of Nuclear and Particle Physics  (France);
Ministry of Education and Science of the Russian Federation, 
National Research Center ``Kurchatov Institute" of the Russian Federation, and 
Russian Foundation for Basic Research  (Russia);
National Council for the Development of Science and Technology and
Carlos Chagas Filho Foundation for the Support of Research in the State of Rio de Janeiro (Brazil);
Department of Atomic Energy and Department of Science and Technology (India);
Administrative Department of Science, Technology and Innovation (Colombia);
National Council of Science and Technology (Mexico);
National Research Foundation of Korea (Korea);
Foundation for Fundamental Research on Matter (The Netherlands);
Science and Technology Facilities Council and The Royal Society (United Kingdom);
Ministry of Education, Youth and Sports (Czech Republic);
Bundesministerium f\"{u}r Bildung und Forschung (Federal Ministry of Education and Research) and 
Deutsche Forschungsgemeinschaft (German Research Foundation) (Germany);
Science Foundation Ireland (Ireland);
Swedish Research Council (Sweden);
China Academy of Sciences and National Natural Science Foundation of China (China);
and
Ministry of Education and Science of Ukraine (Ukraine).


\begin{thebibliography}{99}
\bibitem{d0_coordinate}
The polar angle $\theta$ and the azimuthal angle $\phi$ are defined
with respect to the positive $z$ axis, which is along the proton beam direction.
Pseudorapidity is defined as $\eta=-\ln[\tan(\theta/2)]$.
$\eta_{\rm det}$ and $\phi_{\rm det}$ are the pseudorapidity and the azimuthal angle
measured with respect to the center of the detector.
\end{thebibliography}
\end{document}